\documentclass{ws-ijgmmp}
\usepackage{amssymb}
\usepackage{amsmath}
\usepackage{amssymb}
\usepackage{amsmath}
\usepackage{amsfonts}
\usepackage{amsmath}
\usepackage{graphicx}
\usepackage{amssymb}
\usepackage{color}
\usepackage{enumerate}

\begin{document}

\markboth{Fabio Briscese and Francesco Calogero} {Isochronous
solutions of Einstein's equations and their Newtonian limit}

\catchline{}{}{}{}{}
%
%%%%%%%%%%%%%%%%%%%%%%%%%%%%%%%%%%%%%%%%%%%%%%%%%%%%%%%%%%%%%%%%%%%%

\title{Isochronous solutions of Einstein's equations and their Newtonian
limit}

\author{FABIO BRISCESE}

\address{Department of Physics, Southern University of Science
and Technology, Shenzhen 518055, China\\
and\\
Istituto Nazionale di Alta Matematica Francesco Severi, Gruppo
Nazionale di Fisica Matematica, Citt\`{a} Universitaria, P.le A.
Moro 5, 00185 Rome, Italy.
\\
\email{briscesef@sustech.edu.cn, briscese.phys@gmail.com}}

\author{FRANCESCO CALOGERO}

\address{Dipartimento di Fisica, Universit\`{a} di Roma \textquotedblleft La
Sapienza\textquotedblright , Rome, Italy\\
and\\
Istituto Nazionale di Fisica Nucleare, Sezione di Roma, Rome,
Italy.
\\
\email{francesco.calogero@roma1.infn.it,
francesco.calogero@uniroma1.it} }

\maketitle

\begin{history}
\received{} \revised{}
\end{history}

\begin{abstract}
It has been recently demonstrated that it is possible to construct
isochronous cosmologies, extending to general relativity a result valid for
non-relativistic Hamiltonian systems. In this paper we review these findings
and we discuss the Newtonian limit of these isochronous spacetimes, showing
that it reproduces the analogous findings in the context of non-relativistic
dynamics.
\end{abstract}

\markboth{Fabio Briscese and Francesco Calogero} {Isochronous
solutions of Einstein's equations and their Newtonian limit}

\catchline{}{}{}{}{} %
%%%%%%%%%%%%%%%%%%%%%%%%%%%%%%%%%%%%%%%%%%%%%%%%%%%%%%%%%%%%%%%%%%%%

\address{Department of Physics, Southern University of Science
and Technology, No. 1088, Xueyuan Rd, Xili, Nanshan District,
518055 Shenzhen, Guangdong, P.R. China\\
and\\
Istituto Nazionale di Alta Matematica Francesco Severi, Gruppo
Nazionale di Fisica Matematica, Citt\`{a} Universitaria, P.le A.
Moro 5, 00185 Rome, Italy.
\\
\email{briscese.phys@gmail.com}}

\address{Dipartimento di Fisica, Universit\`{a} di Roma \textquotedblleft La
Sapienza\textquotedblright , Rome, Italy\\
and\\
Istituto Nazionale di Fisica Nucleare, Sezione di Roma, Rome,
Italy.
\\
\email{francesco.calogero@roma1.infn.it,
francesco.calogero@uniroma1.it} }

\begin{history}
\received{} \revised{}
\end{history}

\section{Introduction}

It has been shown \cite{CL2007,calogero} that, for any general autonomous
dynamical system $D$, it is possible to manufacture another autonomous
dynamical system $\tilde{D}$, featuring two additional \textit{arbitrary}
positive parameters $T$ and $\tilde{T}$ with $T>\tilde{T}$ and possibly also
two additional dynamical variables, in such a way that: (1) For the same
variables of the original system $D$ the new system $\tilde{D}$ yields, over
the \textit{arbitrarily long} time interval $\tilde{T}$, a dynamical
evolution which mimics \textit{arbitrarily closely}, or possibly even
\textit{identically}, that yielded by the original system $D$. (2) The
system $\tilde{D}$ is \textit{isochronous}: all its solutions are completely
periodic with an arbitrarily assigned period $T>\tilde{T}$ for any initial
data.

Moreover it has been shown \cite{CL2007,calogero} that, if the dynamical
system $D$ is a many-body problem characterized by a (standard, autonomous)
Hamiltonian $H$ which is translation-invariant (i. e., it features no
external forces), other (also autonomous) Hamiltonians $\tilde{H}$
characterizing \textit{modified} many-body problems can be manufactured
which feature the \textit{same} dynamical variables as $H$ (i. e., in this
case there is no need to introduce two additional dynamical variables) and
which yield a time evolution \textit{arbitrarily close}, or even \textit{%
identical}, to that yielded by the original Hamiltonian $H$ over the \textit{%
\ arbitrarily assigned} time $\tilde{T}$, while being \textit{isochronous}
with the \textit{arbitrarily assigned} period $T$, of course with $T>\tilde{T%
}$.

Let us emphasize that the class of Hamiltonians $H$ for which this result is
valid is quite general. In particular it includes the standard Hamiltonian
system describing an \textit{arbitrary} number $N$ of point particles with
\textit{arbitrary} masses moving in a space of \textit{arbitrary} dimensions
$d$ and interacting among themselves via potentials depending \textit{%
arbitrarily} from the interparticle distances (including the possibility of
multiparticle forces), being therefore generally valid for any realistic
many-body problem, hence encompassing most of non-relativistic physics. This
result is moreover true, \textit{mutatis mutandis}, in a quantal context.

Recently these results, which have been obtained in the context of classical
Hamiltonian mechanics \cite{CL2007,calogero}, have been extended to a
general relativity and cosmological context \cite{isochronous
cosmologies,isochronous spacetimes,isochronous conference paper}. The main
feature of these results is to point out the possibility to provide
alternative descriptions of the relevant physics which are equivalent
\textit{locally} \textit{in time} (for arbitrarily long time intervals) to
the phenomenology being described by the more standard approaches but
predict instead a cyclic behavior (of course over longer time periods).

In our previous paper \cite{isochronous cosmologies} we have exploited this
idea in the context of cosmology, and have shown that, for any
non-degenerate, homogeneous, isotropic and spatially flat metric $g_{\mu \nu
}$ satisfying Einstein's equations and providing a model of the universe at
large scales, it is possible to find a different (also homogeneous,
isotropic and spatially flat but degenerate) metric solution $\tilde{g}_{\mu
\nu }$ which is \textit{locally} (in time) diffeomorphic to $g_{\mu \nu }$
and is periodic with an \textit{arbitrary} period $T>\tilde{T}$ in the time
coordinate $t\,$; but it is \textit{degenerate} at an infinite, discrete
sequence of times $t_{n}=t_{0}\pm nT/2,$ $n=0,1,2,...$ . We interpreted
these metrics as corresponding to \textit{isochronous cosmologies }\cite%
{isochronous cosmologies}.

We emphasize that, due to the diffeomorphic correspondence between $g_{\mu
\nu }$ and $\tilde{g}_{\mu \nu }$ \textit{locally in time}---for time
intervals of order $\tilde{T}$---these two metrics give the same physics
\textit{locally in time}, and therefore there is no way to distinguish them
using observations local in time; which is essentially the same finding
valid in the context of the Hamiltonian systems considered in \cite%
{CL2007,calogero} and tersely recalled above. We stress that, in
the general relativistic context considered in this paper,
\textit{locally in time} refers to time intervals smaller than the
period $T$ of the isochronous metric yet of cosmological scale
(billions of years). Indeed, the statement that the two metrics
$g_{\mu \nu }$ and $\tilde{g}_{\mu \nu }$ give the same physics
locally in time means that this property is valid over time
intervals smaller than, but of the same order as $T$, while any
difference between $g_{\mu \nu }$ and $\tilde{g}_{\mu \nu }$ can
be measured only through observations which last more than $T$.
Obviously, observations over a time period larger than $T$ cannot
be considered \textit{humanly} feasible, so two cosmologies which
can only be distinguished via such experiments are, in the context
of human science, indistinguishable; even though the corresponding
cosmologies may be \textit{de facto} quite different, for instance
one featuring a Big Bang, the other being cyclic but singularity
free for all time.

In fact, while being cyclic\footnote{%
Since periodicity is not an invariant concept (it is not invariant under a
redefinition of time), it is usually preferred to talk of cyclic instead of
periodic solutions of Einstein's equations. For cyclic models in standard
general relativity see \cite{cyclic} .} on time scales larger than $\tilde{T}
$, the metric $\tilde{g}_{\mu \nu }$ may yield over the time interval $%
\tilde{T}$ just the same cosmology which characterize the metric $g_{\mu \nu
}$ of the standard $\Lambda $-CDM cosmological model (see for instance \cite%
{mukhanov} for a review), consistently with all observational tests \cite%
{cmb,leansing,bao,lss,supernovae,ref0,ref1,ref2,ref3,ref4,bicep2}.
Furthermore, it has been shown that $\tilde{g}_{\mu \nu }$ can be
manufactured to be geodesically complete \cite{isochronous
cosmologies,isochronous spacetimes} and therefore singularity-free\footnote{%
A spacetime is singularity-free if it is geodesically complete, i.e. if its
geodesics can be always past- and future-extended \cite{poisson,MTW}.}, so
that the geodesic motion as well as all physical quantities described by
scalar invariants are always well defined, and the Big Bang singularity may
be avoided---even when some of the phenomenological observations can
nevertheless be interpreted as remnants of a past Big Bang, which however
may never be actually attained by the metric $\tilde{g}_{\mu \nu }$, neither
in the past nor in the future.

In \cite{isochronous spacetimes} it has been pointed out that the
realization of the isochronous spacetimes is not restricted to homogeneous,
isotropic and spatially flat metrics, but it can be easily extended to any
synchronous metric. Therefore our result is quite general, since any metric
can be written in synchronous form by a diffeomorphic change of coordinates.
Furthermore, it has been shown that isochronous spacetimes can be realized
in two different ways: the first by means of isochronous metrics that are
\textit{degenerate} at a discrete set of instants $t_{n}$ when the time
reversals occur; the second via non-degenerate metrics featuring a jump in
their first derivatives at the inversion times $t_{n}$, which then implies a
distributional contribution in the stress-energy tensor at $t_{n}$, see the
discussion in \cite{isochronous spacetimes}.

A remark on the terminology: since the geodesics of the isochronous metrics
are spirals \cite{isochronous cosmologies,isochronous spacetimes} of the
form $x(\lambda )=[\lambda ,\vec{V}\left( \lambda \right) ]$ with $\vec{V}%
\left( \lambda \right) $ periodic, we write below that isochronous metrics
correspond to \textit{spiralling} geometries.

Let us clarify that, being \textit{degenerate}, such isochronous
metrics are considered unacceptable in orthodox general
relativity, although they are quite consistent with the
foundational view underlining general relativity that the basis of
cosmology is the \textit{\ geometry} of spacetime rather than the
coordinate system---i. e., the metric---used to describe that
spacetime. In this context a spiraling geometry should be
acceptable; indeed, if it entails no singularity of the metric but
only a degeneracy, it should perhaps be considered more acceptable
than a geometry entailing singularities such as those associated
with the occurrence of a Big Bang and/or a Big Crunch. In any case
the acceptability or not of a
class of metrics should be ultimately decided by experiment, not by \textit{%
a priori} assumptions (for instance, that the metric be
non-degenerate); if such experiments are unfeasible, there is no
justification to exclude \textit{in principle} the consideration
of such metrics. Hence, our point of view that a spiraling
geometry of spacetime should be considered acceptable even if it
requires a relaxation of the requirement that the metric be
locally Minkowskian \textit{everywhere}. On the other hand let us
also note that an alternative description of a spacetime
characterized by a cyclic evolution is also compatible with
non-degenerate metrics, but at the cost of introducing
distributional energy-momentum tensors: see \cite{isochronous
spacetimes}.

In this paper we investigate the weak field (or Newtonian) limit of \textit{%
isochronous spacetimes}. In particular we show that in this limit every
isochronous metric reduces to a solution of an isochronous dynamical system $%
\tilde{D}$ that is obtained from a Newtonian $N$-body system through the
techniques \cite{CL2007,calogero} which have been introduced in the context
of autonomous dynamical systems. In fact, our purpose and scope is to
provide a bridge among the results valid respectively in the general
relativity and in the classical mechanics contexts, by revisiting in the new
context \cite{isochronous cosmologies,isochronous spacetimes,isochronous
conference paper} the standard derivation of Newtonian gravitational physics
from general relativity.

This paper is organized as follows: in section \ref{section review} we
review some mathematical aspects of isochronous metrics, which shall be
useful for the study of the Newtonian limit; this treatment includes an
extended review of previous findings \cite{isochronous
cosmologies,isochronous spacetimes,isochronous conference paper}, which we
considered appropriate given the somewhat "unpleasant"---yet in our opinion
cogent---implications of these findings, which do imply a substantial
limitation of all cosmological theories based on general relativity. In
section \ref{section newtonian limit} we investigate the weak field limit of
our treatment of general relativity and thereby derive the corresponding
Newtonian framework. Finally, we conclude with some general considerations
about our findings, as reported in the last section.

\section{Isochronous solutions of Einstein's equations}

\label{section review}

In this section we review the properties of the isochronous metrics
discussed in \cite{isochronous cosmologies,isochronous
spacetimes,isochronous conference paper}. In what follows Latin indexes such
as $k,h,i,\ldots $ run over the integers from $1$ to $3$ and Greek indexes
such as $\alpha ,\beta ,\gamma ,\ldots $ run over the integers from $0$ to $%
3 $. Let us start from a given metric tensor $g_{\mu \nu }(y)$ in a
coordinate system $y$ defined by
\begin{equation}
ds^{2}=g_{\mu \nu }(y)\,dy^{\mu }dy^{\nu }\,,  \label{metric y}
\end{equation}%
which is a solution of the Einstein's equations%
\begin{equation}
G_{\mu \nu }(y)=\frac{8\pi G}{c^{4}}T_{\mu \nu }(y)\,,
\label{einstein equations y}
\end{equation}%
where $G_{\mu \nu }(y)=R_{\mu \nu }(y)-R(y)\,g_{\mu \nu }(y)/2$ is the
Einstein's tensor constructed with the metric $g_{\mu \nu }(y)$, and $T_{\mu
\nu }(y)$ is the energy-momentum tensor in the reference system $y$.

As in the case of non relativistic Hamiltonian many body systems, the trick
to construct isochronous solutions is based on the introduction of a
"periodic change of time". In the context of general relativity, this is
realized formally by introducing the following "change of coordinates",%
\begin{equation}
dy^{0}=b(x^{0})~dx^{0}~,\qquad dy^{k}=dx^{k}~,  \label{repar}
\end{equation}%
where the function $b(x^{0})$ is periodic with an \textit{\ arbitrarily
assigned} period $T$, $b(x^{0}+T)=b(x^{0})$, having moreover a vanishing
mean value, so that the function $\tau \left( x^{0}\right) $ defined by
\begin{equation}
\tau \left( x^{0}\right) \equiv \int_{0}^{x^{0}}b\left( z\right) dz~
\label{4}
\end{equation}%
is also periodic with period $T$, that is $\tau (x^{0}+T)=\tau (x^{0})$. As
a consequence of these assumptions, the function $b(x^{0})$ must vanish at
an infinite set of time-like hypersurfaces $\Sigma _{n}$ corresponding to
the times $x^{0}=x_{n}^{0}$ where $b(x_{n}^{0})$ changes its sign; thus (\ref%
{repar}) is not a global, but only a local diffeomorphism.

Let us consider a different metric $\tilde{g}_{\mu \nu }(x)$ in the
coordinate system $x$, defined by
\begin{equation}
\begin{array}{ll}
d\tilde{s}^{2}=b(x^{0})^{2}g_{00}\left( y\left( x\right) \right)
\,(dx^{0})^{2}+2\,b(x^{0})\,g_{0k}\left( y\left( x\right) \right)
\,dx^{0}dx^{k}+g_{kh}\left( y\left( x\right) \right) \,dx^{k}dx^{h}\equiv \\
\\
\equiv \tilde{g}_{\mu \nu }(x)\,dx^{\mu }dx^{\nu }\,, \label{metric x} &
\end{array}%
\end{equation}%
where \textbf{$g_{\alpha \beta }\left( y\left( x\right) \right) =g_{\alpha
\beta }\left( y^{0}=\tau (x^{0}),y^{k}=x^{k}\right) $.} Such metric will be
a solution of the equations
\begin{subequations}
\label{einstein equations x}
\begin{equation}
\tilde{G}_{00}(x)=b(x^{0})^{2}\,G_{00}(y(x))=b(x^{0})^{2}\frac{8\pi G}{c^{4}}%
\,T_{00}(y(x))=\frac{8\pi G}{c^{4}}\tilde{T}_{00}(x)\,,
\label{einstein equations x 00}
\end{equation}

\begin{equation}  \label{einstein equations x 0k}
\tilde{G}_{0 k}(x) = b(x^0) \,G_{0 k}(y(x)) = b(x^0)^2 \frac{8 \pi G}{c^4}
\, T_{0 k}(y(x)) = \frac{8 \pi G}{c^4} \tilde{T}_{0 k}(x) \,,
\end{equation}

\begin{equation}
\tilde{G}_{kh}(x)= G_{kh}(y(x)) = \frac{8\pi G}{c^{4}}\,T_{kh}(y(x)) = \frac{%
8\pi G}{c^{4}}\tilde{T}_{kh}(x)\,,  \label{einstein equations x kh}
\end{equation}%
where $\tilde{G}_{\mu \nu }(x)=\tilde{R}_{\mu \nu }(x)-\tilde{R}(x)\,\tilde{g%
}_{\mu \nu }(x)/2$ is the Einstein's tensor constructed with the metric $%
\tilde{g}_{\mu \nu }(x)$, and $\tilde{T}_{\mu \nu }(x)$ is defined by (\ref%
{einstein equations x}). From (\ref{einstein equations x}) it is evident
that the metric (\ref{metric x}) satisfies Einstein's equations everywhere
except at the hypersurfaces $\Sigma _{n}$, where it is degenerate, and where
all the quantities containing the inverse tensor $\tilde{g}^{\mu \nu }$, as
the affine connection $\tilde{\Gamma} _{\,\,\,\,\beta \gamma }^{\alpha }$
and the Ricci and Riemann curvature tensors, are \textit{not} defined.
However we note that, even if it is degenerate on the hypersurfaces $\Sigma
_{n}$, the curvature tensor $\tilde{g}_{\mu \nu }(x)$ is continuous and
differentiable at $\Sigma _{n}$, provided $g_{\mu \nu }(y)$ is
differentiable on the hypersurfaces $\tilde{\Sigma}_{n}$ defined by $%
y^{0}=y_{n}^{0}\equiv \tau (x_{n})$. Moreover, the scalar curvature $\tilde{R%
}(x)=R(y(x))$ is also continuous on $\Sigma _{n}$, provided $R(y)$ is
regular on $\tilde{\Sigma}_{n}$. This resembles the divergence of the
Schwarzschild metric on the Schwarzschild horizon, where the metric tensor
diverges but the Ricci scalar curvature $\tilde{R}$ remains finite, and the
singularity in the metric tensor does not imply a physical singularity on
the horizon. However, the situation here is different in the sense that the
metric $\tilde{g}_{\mu \nu }(x)$ is differentiable but degenerate on $\Sigma
_{n}$, and it has to be interpreted in the framework of degenerate solutions
of Einstein's equations (see the discussion in \cite{isochronous spacetimes}%
). Indeed, the metric (\ref{metric x}) does not entail any physical
singularity on $\Sigma _{n}$, since all the physical quantities (such as,
for instance, the Ricci scalar $\tilde{R}$) remain finite there, but it
corresponds to a metric tensor which is periodic in time and becomes
degenerate at the infinite set $\Sigma _{n}$. The degeneracy of the metric
tensor implies the failure of the equivalence principle on $\Sigma _{n}$, in
its formulation that states that the signature of the metric tensor must be
Minkowskian. However, it is a matter of discussion whether this fact could
have observable effects in an isochronous world, where the dynamics of all
the universe, including that of its components, is periodic.

Degenerate metrics have been studied in the past, motivated by the
consideration of the so called signature-changing metrics \cite%
{ellis1,ellis2,elliscarfora,elliscomment,ellis3,ellis4,ellis5,ellis6,ellis8,ellis9,ellis10,ellis11,ellis12,ellis13,ellis14,ellis15,ellis17,ellis18,ellis19}%
, which give a classical realization of the change of signature in quantum
cosmology conjectured by Hartle and Hawking \cite%
{hartlehawking3,hartlehawking1,hartlehawking2,hartlehawking4,hawkinghistory}%
. The classical change of signature for a homogeneous, isotropic and
spatially flat universe is realized by a metric tensor defined by the
following line element,
\end{subequations}
\begin{equation}
d\tilde{s}^{2}=N(t)~dt^{2}-\,a(t)^{2}~d\vec{x}^{2}~,
\label{changeofsignaturedegenerate}
\end{equation}%
the lapse function $N(t)$ being a continuous function changing sign at some
time $t_{0}$
%, for instance $N(t)$ is positive for $t>t_{0}$, negative for $t<t_{0}$ and zero for $t=t_{0}$
where the metric (\ref{changeofsignaturedegenerate}) is degenerate. Note
that this situation is physically different for that of the metric tensor (%
\ref{metric x}), where the lapse function $N(t)=b^{2}\left( t\right) $
vanishes on $\Sigma _{n}$, but it never becomes negative.

The inclusion of degenerate metrics as (\ref{metric x}) and (\ref%
{changeofsignaturedegenerate}) in the context of general relativity does not
by itself entail well defined physical implications, which also depend on
the prescriptions defining the behavior at the points of degeneracy of the
metric: \textit{different} prescriptions yield \textit{different} physical
theories \cite{elliscomment}. In particular different generalizations based
on different prescriptions give different junction conditions on the
degeneracy hypersurfaces, see for instance the discussion in \cite%
{elliscarfora,elliscomment}. In fact "one can also consider discontinuities
in the extrinsic curvature" \cite{elliscarfora} on the degeneracy
hypersurface of a degenerate solution of Einstein's equations, "but attempts
to relate these to a distributional matter source at the boundary require
some form of field equations valid on the surface" \cite{elliscomment}, and
this would require some special prescription on what Einstein's equations
are on the degeneracy hypersurfaces. Thus, if one accepts degenerate
solutions of Einstein's equations as physically acceptable spacetimes, there
is no reason to infer that a discontinuity of the extrinsic curvature on a
degeneracy surface implies that the stress energy tensor associated to such
a metric tensor must have a distributional form there.

One might object that the isochronous metric (\ref{metric x}) entails a
discontinuity of the extrinsic curvature $K_{ab}$ on the hypersurfaces $%
\Sigma _{n}$, which, in the context of Einstein's theory, means that the
energy-momentum tensor has distributional character on $\Sigma _{n}$. The
point here is that the construction which implies this relation is no longer
valid for isochronous metrics of the type (\ref{metric x}), since the
unitary normal vectors to the hypersurfaces $\Sigma _{n}$ do not exist
because the metric is degenerate on $\Sigma _{n}$.

To discuss this point, let us recall the Darmois formalism \cite{ellis16},
which defines the junction condition under which two different metric
tensors can be joined on some hypersurface. Let us consider a hypersurface $%
\Sigma $ dividing the spacetime in two regions $V^{\left( +\right) }$ and $%
V^{\left( -\right) }$. The condition that the two metrics $g_{\mu \nu
}^{\left( +\right) }$ and $g_{\mu \nu }^{\left( -\right) }$ in the two
regions $V^{\left( +\right) }$ and $V^{\left( -\right) }$ must satisfy in
order to join smoothly on $\Sigma $ is that they must be the same on both
sides of $\Sigma $ together with their first derivatives, see for instance
Eq.(3.7.7) in \cite{poisson}, that is
\begin{subequations}
\label{junction 0}
\begin{equation}
g_{\mu \nu }^{\left( +\right) }|_{\Sigma }=g_{\mu \nu }^{\left( -\right)
}|_{\Sigma }~,  \label{junction 0 a}
\end{equation}

\begin{equation}
g_{\mu \nu ,\sigma }^{\left( +\right) }|_{\Sigma } = g_{\mu \nu ,\sigma
}^{\left( -\right) }|_{\Sigma }~.
\end{equation}

From (\ref{junction 0}) it is therefore evident that the isochronous metric (%
\ref{metric x}) satisfies such junction conditions on the hypersurfaces $%
\Sigma_n$, where it is infinitely differentiable.

In Einstein's gravity (which implies a restriction to non-degenerate
metrics) such a condition is usually expressed in an invariant way by use of
the extrinsic curvature, which is diffeomorphism-invariant. Actually the
second equation in (\ref{junction 0}) is expressed in terms of the
derivatives $g_{\mu \nu ,\sigma }$ of the metric tensor, which are not
themselves tensors. Therefore (\ref{junction 0}) is more conveniently recast
in tensor form as
\end{subequations}
\begin{equation}
g_{\mu \nu }^{\left( +\right) }|_{\Sigma }=g_{\mu \nu }^{\left( -\right)
}|_{\Sigma }~,\qquad K_{ab}^{\left( +\right) }|_{\Sigma }=K_{ab}^{\left(
-\right) }|_{\Sigma }~,  \label{junction}
\end{equation}%
where $K_{ab}$ is the extrinsic curvature of $\Sigma $ defined as
\begin{equation}
K_{ab}=n_{\alpha ;\beta }\,\,e_{a}^{\alpha }\,\,e_{b}^{\beta }~,
\label{extrinsic curvature 0}
\end{equation}%
with $n^{\alpha }$ the unitary normal vector to $\Sigma $ and $e_{a}^{\alpha
}$ three unitary tangent vectors to $\Sigma $, so that $g_{\alpha \beta
}=n_{\alpha }n_{\beta }-e_{\alpha a}e_{\beta b}\delta ^{ab}$, where $%
n_{\alpha }=g_{\alpha \beta }\,n^{\beta }$ and $e_{\alpha a}=g_{\alpha \beta
}\,e_{a}^{\beta }$, and $g_{|_{\Sigma }\,ab}\equiv g_{\alpha \beta
}\,e_{a}^{\alpha }\,e_{b}^{\beta }$ is the induced metric over $\Sigma $.
Note that (\ref{junction 0}) is more general than (\ref{junction}), since it
is valid even where the extrinsic curvature does not exist. In this context
a discontinuity of the extrinsic curvature on the hypersurface $\Sigma $ is
associated with a singularity of the stress-energy tensor, which turns out
to have distributional character (corresponding to the presence of thin
shells) on $\Sigma $, acquiring a distributional contribution given by
\begin{equation}
T_{\alpha \beta }^{distr}=\delta _{\Sigma }\,\,S_{ab}|_{\Sigma }e_{\alpha
}^{a}e_{\beta }^{b}~,  \label{StressEnergyTensorNondegenerate}
\end{equation}%
where $\delta _{\Sigma }$ is the delta function with support on the
hypersurface $\Sigma $, and
\begin{equation}
S_{ab}|_{\Sigma }\equiv \frac{1}{8\pi }\left( [K_{ab}]|_{\Sigma
}-[K]|_{\Sigma }\,\,\tilde{g}_{|_{\Sigma }\,ab}\right),
\end{equation}
with
\begin{equation}
[K_{ab}]|_{\Sigma }\equiv K_{ab}|_{\Sigma ^{+}}-K_{ab}|_{\Sigma ^{-}}, \quad
\lbrack K]|_{\Sigma }\equiv [K_{ab}]|_{\Sigma }\,\,g_{|_{\Sigma }}^{ab}\,.
\end{equation}

However, this result is valid in Einstein's theory, which assumes that the
metric tensor has Minkowskian signature, and therefore is non-degenerate.
The case of the isochronous metric $\tilde{g}$ given in (\ref{metric x}) is
different, since $\tilde{g}$ is differentiable on $\Sigma _{n}$ and
therefore it satisfies the junction conditions in the form (\ref{junction 0}%
); however, since $n^{\alpha }$ is not defined on $\Sigma _{n}$, the
junction conditions cannot be recast in the form (\ref{junction}).
Therefore, for degenerate metrics, the formalism which leads to (\ref%
{StressEnergyTensorNondegenerate}) does not apply, and one is not allowed to
conclude that a discontinuity of the extrinsic curvature implies that the
stress energy tensor has distributional character on $\Sigma _{n}$.

What is more, delta functions with support on degeneracy hypersurfaces have
no significant effect in covariant integrals \cite{ellis15,ellis19}, due to
the fact that the covariant volume element $dx^{4}\sqrt{-\tilde{g}}$
associated to a degenerate metric is null on its degeneracy hypersurface,
because the determinant $\tilde{g}$ vanishes there. For instance, for the
metric (\ref{metric x}) one has that $\int dx^{4}\sqrt{-\tilde{g}}\delta
(t-t_{n})f\left( x\right) =0 $, for any integrable function $f\left(x\right)$%
. Hence delta-like distributions centered on the degeneracy hypersurfaces $%
t=t_{n}$ disappear from integrals (see also the discussion in \cite{ellis15}%
).

So, the question of how to include degenerate metrics in general relativity
is moot. In order to formulate a generalized gravitational theory which
includes such metrics, it is convenient to focus directly on the Einstein's
equations characterizing the geometry of spacetime, which themselves admit
degenerate metrics as their solutions. If, however, one considers desirable
to derive this theory from a variational principle, a possibility is to
assume the standard Einstein-Hilbert gravitational action of general
relativity, such that the total action of gravitation plus the matter
(nongravitational) fields is%
\begin{equation}
S_{Tot}=\int_{M}\sqrt{-\tilde{g}}\left( -\frac{1}{2k}\tilde{R}%
+L_{Mat}\right) d^{4}x~,  \label{actiontotal}
\end{equation}%
where $k=8\pi G/c^{4}$, $G$ is the gravitational constant, $c$ the speed of
light, $M$ is the manifold defining the spacetime and $L_{Mat}$ is the
Lagrangian density of matter fields; and to add the requirement that, for a
metric tensor which is degenerate on some hypersurface $\Sigma $, the
variation of the metric $\tilde{g}_{\mu \nu }$ vanish on $\Sigma $, i.e. $%
\delta \tilde{g}_{\mu \nu }|_{\Sigma }=0$. With such assumption the
variation of the action (\ref{actiontotal}) is%
\begin{equation}
\delta S_{Tot}=\frac{1}{2}\int_{M/\Sigma }\sqrt{-\tilde{g}}\left( -\frac{1}{k%
}\tilde{G}^{\mu \nu }+\tilde{T}^{\mu \nu }\right) d^{4}x~,
\label{actiontotalvariation}
\end{equation}%
where $\tilde{G}^{\mu \nu }$ is the Einstein tensor and $\tilde{T}^{\mu \nu
} $ is the stress-energy tensor of matter fields, which gives the standard
equations of general relativity in the region where the metric tensor is
non-degenerate.

To bypass the problems related to degenerate signature-changing metrics, it
has been proposed \cite{ellis1,ellis2} to consider a different,
non-degenerate but discontinuous, realization of the classical change of
signature, as given by the metric%
\begin{equation}
ds^{2}=f(\tau )~d\tau ^{2}-\, \alpha(\tau )^{2}~d\vec{x}^{2},
\label{changeofsignaturediscontinuous}
\end{equation}%
with a discontinuous lapse function $f(\tau )=1$ for $\tau >\tau _{0}$ and $%
f(\tau )=-1$ for $\tau <\tau _{0}$. It has been shown \cite{ellis1,ellis2}
that in this case it is possible to introduce a smooth (generalized)
orthonormal reference frame which allows a variational derivation of
Einstein's equations. In such case, the Darmois formalism \cite{ellis16}
applies, in such a way that the discontinuity of the extrinsic curvature on
a hypersurface is related to a distributional stress energy tensor \cite%
{ellis15}. This is achieved by adding a surface term to the Einstein-Hilbert
action, in such a way that the gravitational action becomes%
\begin{equation}
S_{g}=\int_{M/\Sigma }\sqrt{- g}\,R \,d^{4}x+\oint_{\Sigma }\sqrt{- g } \, K
\, d\Sigma \, ,  \label{actionboundary}
\end{equation}%
where $M$ is the manifold defining the spacetime, $\Sigma $ is the boundary
of $M$ where the metric tensor is discontinuous, $R$ is the Ricci scalar
curvature and $K$ the extrinsic curvature of $\Sigma $. The addition of such
a boundary term is always necessary when one considers manifolds with
boundaries \cite{HawkingHorowitz}. Incidentally we note that this action
cannot be used in the case of degenerate metrics, because in this case the
degeneracy surface $\Sigma $ has no unitary normal vector hence the
extrinsic curvature $K$ does not exist.

The two realizations of the classical change of signature, the continuous
and degenerate one, see (\ref{changeofsignaturedegenerate}), and the
discontinuous one, see (\ref{changeofsignaturediscontinuous}), are \textit{%
locally but not globally} diffeomorphic, since they are related by a change
of the time variable $dt=d\tau /\sqrt{|N(t)|}$ which is not defined at the
time of signature change $t_{0}$ when $N(t)=0$. Hence they represent two
different spacetimes. However, except for the instants when $N(t)$ vanishes,
they are \textit{locally } diffeomorphic and thus they describe---excepts at
those instants---the same physics.

As in the case of signature-changing metrics, it is also possible to
consider a realization of \textit{isochronous} cosmologies different from (%
\ref{metric x}), via \textit{non-degenerate} continuous metrics featuring a
finite jump of their first derivatives at an infinite, discrete set of
equispaced times. Such a realization is given by the formal change of time $%
d\hat\tau =b(x^{0})dx^{0}$ in (\ref{metric x}), corresponding to $%
dy^{0}=C(\hat\tau )~d\hat\tau $ in (\ref{metric y}), with $C(\hat\tau )=1$
for $2nT<\hat\tau <(2n+1)T/2$ and $C(\hat\tau )=-1$ for $(2n+1)T/2<\hat\tau
<2(n+1)T$ and $n$ integer, giving%
\begin{equation}
d\tilde{s}^{2}=g_{00}\left( y\left( \hat\tau \right) \right) \,(d\hat\tau
)^{2}+2g_{0k}\left( y\left( \hat\tau \right) \right) \,d\hat\tau
dy^{k}+g_{kh}\left( y\left( \hat\tau \right) \right) \,dy^{k}dy^{h}\equiv
\tilde{g}_{\mu \nu }(x)\,dx^{\mu }dx^{\nu }\,.  \label{metric eta}
\end{equation}%
The metric (\ref{metric eta}) is isochronous and it is continuous but not
differentiable at the set of times $\hat\tau _{n}=nT$, and it entails a
discontinuity of the extrinsic curvature at this infinite set of times.
Since, as in the case of (\ref{changeofsignaturediscontinuous}), it is
possible to introduce a smooth (generalized) orthonormal reference frame for
(\ref{metric eta}), the Darmois formalism \cite{ellis16} applies and
therefore one concludes that the discontinuity of the extrinsic curvature on
$\hat\tau _{n}$ is related to a distributional stress energy tensor \cite%
{ellis15}.

At this point, we must emphasize that the two metrics (\ref{metric x}) and (%
\ref{metric eta}) are two \textit{different} realizations of an \textit{%
isochronous} spacetime, since they are \textit{locally}, but \textit{not
globally}, diffeomorphic via the change of variable $d\hat{\tau}%
=b(x^{0})dx^{0}$, which is singular at $\Sigma _{n}$. In particular, both (%
\ref{metric x}) and (\ref{metric eta}) are \textit{locally} (but \textit{not
globally}) diffeomorphic to the Einsteinian metric (\ref{metric y}), and
therefore they give an \textit{identical} dynamics in any region of
spacetime where $x^{0}\neq x_{n}^{0}$ or $\hat{\tau}\neq \hat{\tau}_{n}$.
The main difference is that the isochronous evolution given by (\ref{metric
x}) entails the introduction of degenerate metrics and the failure of the
equivalence principle at $\Sigma _{n}$, and the non-degenerate realization
of the isochronous dynamics given by (\ref{metric eta}) implies the
existence of a distributional contribution to the stress-energy tensor. In
both cases the evolution of the universe is locally the same as that
described by Einstein's equations, implying that it is not possible to
discriminate by experiments \textit{local in time} between a cyclic and a
noncyclic evolution of the universe.

A last consideration concerns the geodesic completeness of the isochronous
metric (\ref{metric x}). In fact isochronous spacetimes can be manufactured
to be geodesically complete, that is singularity free. This happens for
instance if the Einsteinian metric (\ref{metric y}) is singularity free for $%
y^{0}\in \mathcal{Y}^{0}$, where the interval $\mathcal{Y}^{0}$ is defined
as the image of the real line $\mathbb{R}$ through the function $\tau
(x^{0}) $, in formulas $\mathcal{Y}^{0}:\left\{ y^{0}=\tau \left(
x^{0}\right) \right\} ,\forall x^{0}\in \mathbb{R}$. An example of
singularity free isochronous metric has been discussed in \cite{isochronous
cosmologies,isochronous spacetimes,isochronous conference paper}, obtaining
an isochronous realization of the homogeneous and spatially flat
Friedmann-Robertson-Walker metric with no big bang singularity.

To derive the properties of the geodesics of the metric (\ref{metric x}) one
simply notes that these geodesics can be obtained from those of (\ref{metric
y}). In fact the geodesic equation for (\ref{metric y}) is
\begin{equation}
\frac{d^{2}y^{\beta }}{d\lambda ^{2}}~g_{\beta \alpha }+\Gamma_{\alpha \beta
\gamma }~\frac{dy^{\beta }}{d\lambda }\frac{dy^{\gamma }}{d\lambda }%
=g_{\beta \alpha }~\frac{dy^{\beta }}{d\lambda }\frac{1}{L}\frac{d{L}}{%
d\lambda }~,  \label{geod y}
\end{equation}%
where $\Gamma_{\alpha \beta \gamma }$ are the Christoffel symbols
constructed with the metric $g_{\mu \nu }$, $\lambda $ is a parameter used
to parameterize the geodesic and $L=\sqrt{(dy^{\alpha }/d\lambda )(dy^{\beta
}/d\lambda )g_{\alpha \beta }}$ for timelike geodesics, $L=\sqrt{%
-(dy^{\alpha }/d\lambda )(dy^{\beta }/d\lambda )g_{\alpha \beta }}$ for
spacelike geodesics (an analogous treatment applies to null geodesics).

In the same way, the geodesics of (\ref{metric x}) satisfy the equation%
\begin{equation}
\frac{d^{2}x^{\beta }}{d\lambda ^{2}}~\tilde{g}_{\beta \alpha }+\tilde{\Gamma%
} _{\alpha \beta \gamma }~\frac{dx^{\beta }}{d\lambda }\frac{dx^{\gamma }}{%
d\lambda } =\tilde{g}_{\beta \alpha }~\frac{dx^{\beta }}{d\lambda }\frac{1}{%
\tilde{L}}\frac{d{\tilde{L}}}{ d\lambda }~,  \label{geod
x}
\end{equation}%
where again $\tilde{\Gamma} _{\alpha \beta \gamma }$ are the Christoffel
symbols constructed with $\tilde{g}_{\mu \nu }$, $\lambda $ is a parameter
and $\tilde{L}=\sqrt{ (dx^{\alpha }/d\lambda )(dx^{\beta }/d\lambda )\tilde{g%
}_{\alpha \beta }}$ for timelike geodesics, $\tilde{L}=\sqrt{-(dx^{\alpha
}/d\lambda )(dx^{\beta }/d\lambda )\tilde{g}_{\alpha \beta }}$ for spacelike
geodesics.

Note that (\ref{geod x}) is adequate to define the geodesics even where the
metric $\tilde{g}_{\alpha \beta }$ is not invertible. Also note that we have
not chosen $d\lambda $ to coincide with the line element $d\tilde{s}$, since
in the case of the isochronous metric (\ref{metric x}) the line element
vanishes at $\Sigma _{n}$ and therefore is not a good parameter for the
geodesics. Moreover, when $\lambda $ is not identified with $s$, the
equations (\ref{geod y}-\ref{geod x}) for $\alpha =0$ reduce to an identity;
indeed the functions $y^{0}(\lambda )$ and $x^{0}(\lambda )$ can be chosen
arbitrarily. This arbitrariness reflects the fact that the action
\begin{equation}
\mathcal{L}=\int \sqrt{\tilde{g}_{\alpha \beta }\frac{dx^{\alpha }}{d\lambda
}\frac{dx^{\beta }}{d\lambda }}d\lambda \,,  \label{action}
\end{equation}%
from which the geodesic equation is deduced, is invariant under the
reparameterization $\lambda \rightarrow f(\lambda )$, and therefore the
arbitrariness in the choice of $\lambda $ implies the arbitrariness in the
choice of $y^{0}(\lambda )$ and $x^{0}(\lambda )$. However, when $\lambda $
is forced to coincide with $s$, the functional form of $y^{0}(\lambda )$ and
$x^{0}(\lambda )$ is fixed. We will use this property of the geodesic
equation in the next section.

It is straightforward to show that the geodesics of (\ref{metric x}) are
given by
\begin{equation}
x(\lambda ) = \left[ \lambda,~\vec{y}\left( \tau (\lambda)\right) \right] ~,
\label{geodesics
x}
\end{equation}%
provided that the geodesics of (\ref{metric y}) have the expression
\begin{equation}
y(\lambda )=\left[ \lambda ,~\vec{y}\left( \lambda \right) \right] ~.
\label{geodesics y}
\end{equation}%
It is therefore evident that if $\vec{y}\left( \tau (t)\right) $ is
nonsingular, the geodesics (\ref{geodesics x}) are always definite, and (\ref%
{metric x}) is geodesically complete, thus singularity free, and this
happens if the region of the spacetime (\ref{metric y}) such that $y^{0}\in
\mathcal{Y}^{0}$ is singularity free.

As we have already emphasized in the introduction, since $\tau (\lambda )$
is periodic the geodesics (\ref{geodesics x}) are spirals, and we refer to
this feature when we say that isochronous metrics correspond to spiralling
spacetimes.

\section{Newtonian limit of the isochronous solutions}

\label{section newtonian limit}

Let us start this section by discussing the Newtonian limit for a system of
point-like bodies in the framework of general relativity, see \cite%
{newtonian limit} for a review of the Newtonian limit in general relativity.

Let us first consider the metric (\ref{metric y}). We make the hypothesis
that the gravitational field is weak and the bodies under consideration are
moving slowly. Moreover, we assume that the metric tensor is almost
Minkowskian, so that
\begin{equation}
g_{\mu \nu }(y)=\eta _{\mu \nu }+h_{\mu \nu }(y)~,
\end{equation}%
where $\eta _{\mu \nu }$ is the Minkowski metric, and the weak field and
slow motion approximations read respectively
\begin{equation}
|h_{\mu \nu }|\ll 1\,\qquad |\frac{dy_{i}^{k}}{d\lambda _{i}}|\ll |\frac{%
dy_{i}^{0}}{d\lambda _{i}}|,\quad k=1,2,3~.
\label{weak field slow motion appr}
\end{equation}%
Solving the linearized Einstein's equations one has \cite{newtonian limit}
\begin{equation}
h_{00}{\left( y\right) }\simeq -2~G~\sum_{i}\left[ m_{i}~\frac{\theta \left(
\left\vert \vec{y}-\vec{y}_{i}\right\vert \right) }{\left\vert \vec{y}-\vec{y%
}_{i}\right\vert }\right] ~,  \label{h00 explicit}
\end{equation}%
where%
\begin{equation}
y_{i}(\lambda _{i})=\left[ y_{i}^{0}\left( \lambda _{i}\right) ,\vec{y}%
_{i}\left( \lambda _{i}\right) \right]
\end{equation}%
is the geodesic trajectory of the $i$-th point-like particle of mass $m_{i}$%
, parameterized as a function of the (unphysical) parameter $\lambda _{i}$,
and $\theta $ is a "smoothened" version of the step function such that $%
\theta (z)\simeq 1$ for $z>0$ and $\theta (z)\simeq 0$ for $z<0$ and $%
\lim_{z\rightarrow 0^{+}}\theta (z)/z=0$.

The relation between $h_{00}$ and the Newtonian gravitational potential $%
\phi \left( y\right) $ is $h_{00}\left( y\right) =2\phi \left( y\right)$
and, due to the limit property of $\theta(z)$, when evaluated along the
trajectory $y_i(\lambda)$ the Newtonian potential is
\begin{equation}
\phi {\left( y_{i}\right) }=-~G~\sum_{k\neq i}\left[ m_{i}~\frac{1}{%
\left\vert \vec{y_{i}}-\vec{y}_{k}\right\vert }\right] ~.
\label{h00 explicit}
\end{equation}

The geodesic equations of motion (\ref{geod y}) for the point-like objects
of mass $m_{i}$ and trajectory $y_{i}(\lambda _{i})$ read
\begin{equation}
\frac{d^{2}y_{i}^{\beta }}{d\lambda _{i}^{2}}~g_{\beta \alpha
}(y_{i})+\Gamma_{\alpha \beta \gamma }(y_{i})~\frac{dy_{i}^{\beta }}{%
d\lambda _{i}}\frac{dy_{i}^{\gamma }}{d\lambda _{i}}=g_{\beta \alpha
}(y_{i})~\frac{dy_{i}^{\beta }}{d\lambda _{i}}\frac{1}{L(y_{i})} \frac{d{%
L(y_{i})}}{d\lambda _{i}}~,  \label{geod1 y}
\end{equation}%
where $L(y_{i})=\sqrt{\frac{dy_{i}^{\alpha }}{d\lambda _{i}}\frac{%
dy_{i}^{\beta }}{d\lambda _{i}}g_{\alpha \beta }(y_{i})}$ since we are
considering massive objects, i.e. space-like geodesics.

In the weak field and the slow motion approximations, and assuming that the
gravitational field is slowly varying in time, so that $\Gamma_{000} =
h_{00,0}/2 \simeq 0$ and $\Gamma_{i00} = h_{0i,0} -h_{00,1}/2 \simeq
-h_{00,1}/2$, from (\ref{geod1 y}) one has

\begin{equation}
\frac{d^{2}~y_{i}^{0}}{d~\lambda _{i}^{2}}+\Gamma _{000}\left( \frac{%
d~y_{i}^{0}}{d~\lambda _{i}}\right) ^{2}\simeq \frac{d^{2}~y_{i}^{0}}{%
d~\lambda _{i}^{2}}=\frac{1}{L(y_{i})}\frac{d{L(y_{i})}}{d\lambda _{i}}~,
\label{newtonian limit y 1}
\end{equation}

\begin{equation}
-\frac{d^{2}~y_{i}^{k}}{d~\lambda _{i}^{2}}+\Gamma _{i00}\left( \frac{%
d~y_{i}^{0}}{d~\lambda _{i}}\right) ^{2}=-\frac{d^{2}~y_{i}^{k}}{d~\lambda
_{i}^{2}}-\frac{1}{2}~\frac{\partial h_{00}(y_{i})}{\partial y_{i}^{k}}%
\left( \frac{d~y_{i}^{0}}{d~\lambda _{i}}\right) ^{2}=-\frac{d~y_{i}^{k}}{%
d~\lambda _{i}}\frac{1}{L(y_{i})}\frac{d{L(y_{i})}}{d\lambda _{i}}~,
\label{newtonian limit y 2}
\end{equation}%
where, above and hereafter, $k=1,2,3$.

Using (\ref{weak field slow motion appr}) one has
\begin{equation}
L(y_{i})\simeq \left\vert \frac{d~y_{i}^{0}}{d~\lambda _{i}}\right\vert
\qquad \Rightarrow \qquad \frac{1}{L(y_{i})}\frac{d{L(y_{i})}}{d\lambda _{i}}%
\simeq \left\vert \frac{d~y_{i}^{0}}{d~\lambda _{i}}\right\vert ^{-1}\frac{%
d\left\vert \frac{d~y_{i}^{0}}{d~\lambda _{i}}\right\vert }{d{\lambda _{i}}}%
=\left( \frac{d~y_{i}^{0}}{d~\lambda _{i}}\right) ^{-1}\frac{d^{2}~y_{i}^{0}%
}{d~\lambda _{i}^{2}}~,
\end{equation}%
so that (\ref{newtonian limit y 1}) becomes an identity, hence the
functional form of $y_{i}^{0}(\lambda _{i})$ is not fixed. As we have
already noted, this is due to the arbitrariness of the action (\ref{action})
under the reparameterization $\lambda \rightarrow f(\lambda )$. Furthermore,
(\ref{newtonian limit y 2}) becomes

\begin{equation}  \label{newtonian limit y 3}
\frac{d^{2}~y_i^{k}}{d~\lambda_i ^{2}} - \frac{d^2~y_{i}^{0}}{d~\lambda^2
_{i}} \, \frac{\frac{d~y_{i}^{k}}{d~\lambda _{i}} }{\frac{d~y_{i}^{0}}{%
d~\lambda _{i}}} = - \frac{1}{2} ~\frac{\partial h_{00}(y_i)}{\partial
y_i^{k}} \left( \frac{d~y_i^{0}}{d~\lambda_i } \right)^2.
\end{equation}

Since the functional form of $y_{i}^{0}(\lambda _{i})$ is not fixed by the
dynamics, being $\lambda _{i}$ an unphysical parameter, one has the freedom
to choose $y_{i}^{0}=\lambda _{i}$, so that (\ref{newtonian limit y 3})
reduces to

\begin{equation}
\frac{d^{2}y_{i}^{k}(y_{i}^{0})}{dy_{i}^{0\,2}}=-\frac{1}{2}~\frac{\partial
h_{00}(y_{i})}{\partial y_{i}^{k}}=-\left[ \nabla _{y_{i}}~\phi \left(
y_{i}\right) \right] ^{(k)}~,  \label{newton equation y}
\end{equation}%
where the Newtonian potential $\phi \left( y_{i}\right) $ is given by
\begin{equation}
\phi \left( y_{i}\right) =\frac{1}{2}h_{00}{\left( y_{i}\right) }%
=-~G~\sum_{k\neq i}\left[ m_{i}~\frac{1}{\left\vert \vec{y_{i}}-\vec{y}%
_{k}\right\vert }\right] ~.  \label{newton equation phi}
\end{equation}

Finally, since time is absolute in Newtonian gravity, we set $%
y_{i}^{0}=\lambda _{i}=t$ in (\ref{newton equation y}) for all the
point-like masses $m_{i}$, obtaining Newton's equations
\begin{equation}
\frac{d^{2}y_{i}^{k}(t)}{dt^{2}}=-\partial _{y_{i}^{k}}\phi \left(
y_{i}\right) \,.
\end{equation}

At this point we can analyze the Newtonian limit in the corresponding
isochronous metric (\ref{metric x}). Following our prescription (\ref{repar}%
), we can write the isochronous realization of the weak field and slow
motion limit as
\begin{equation}
\tilde{g}_{00}=b^{2}\left( x_{0}\right) ~\left( \eta _{00}+h_{00}\right)
~,\qquad \tilde{g}_{0i}=b\left( x_{0}\right) ~\left( \eta
_{0i}+h_{0i}\right) ~,\qquad \tilde{g}_{ij}=\eta _{ij}+h_{ij}~\,,
\label{metric x newtonian limit}
\end{equation}%
where $h_{\alpha \beta }=h_{\alpha \beta }\left( y_{i}(x_{i})\right) $.

The geodesic equations (\ref{geod x}) for the \textit{i}-th point-like mass
read%
\begin{equation}
\frac{d^{2}x_{i}^{\beta }}{d\lambda _{i}^{2}}~\tilde{g}_{\beta \alpha
}(x_{i})+{\ \tilde{\Gamma} }_{\alpha \beta \gamma }(x_{i})~\frac{%
dx_{i}^{\beta }}{ d\lambda _{i}}\frac{dx_{i}^{\gamma }}{d\lambda _{i}}=
\tilde{g}_{\beta \alpha }(x_{i})~\frac{dx_{i}^{\beta }}{d\lambda _{i}}\frac{1%
}{\tilde{L}(x_{i})} \frac{d{\tilde{L}(x_{i})}}{d\lambda _{i}}~,
\label{newtonian
limit x 1}
\end{equation}%
where $\tilde{L}(x_{i})=\sqrt{\frac{dx_{i}^{\alpha }}{d\lambda _{i}}\frac{%
dx_{i}^{\beta }}{d\lambda _{i}}\tilde{g}_{\alpha \beta }(y_{i})}$ for
massive point-like objects. More explicitly, the geodesic equations (\ref%
{geod x}) for the space and time components $x^{\alpha }$ are%
\begin{equation}
\begin{array}{ll}
\tilde{g}_{00}\frac{d^{2}~x_{i}^{0}}{d~\lambda _{i}^{2}}+\tilde{\Gamma}
_{000}\left( \frac{d~x_{i}^{0}}{d~\lambda _{i}}\right) ^{2}=b^{2}\left(
x_{i}^{0}\right) ~\left( \frac{d^{2}~x_{i}^{0}}{d~\lambda _{i}^{2}}+\frac{
b^{\prime }\left( x_{i}^{0}\right) }{b\left( x_{i}^{0}\right) }~\left( \frac{
d~x_{i}^{0}}{d~\lambda _{i}}\right) ^{2}\right) = \\
\\
=b^{2}\left( x_{i}^{0}\right) ~\frac{d~x_{i}^{0}}{d~\lambda _{i}}\frac{1}{%
\tilde{L}(x_{i}) }\frac{d{\tilde{L}(x_{i})}}{d\lambda _{i}} \label{newtonian
limit x 2} &
\end{array}%
\end{equation}

and%
\begin{equation}
\begin{array}{ll}
\tilde{g}_{kj}\frac{d^{2}~x_{i}^{j}}{d~\lambda _{i}^{2}}+\tilde{\Gamma}%
_{k00}\left( \frac{d~x_{i}^{0}}{d~\lambda _{i}}\right) ^{2}=-\frac{%
d^{2}~x_{i}^{k}}{d~\lambda _{i}^{2}}+b^{2}\left( x_{i}^{0}\right) ~\left(
\frac{d~x_{i}^{0}}{d~\lambda _{i}}\right) ^{2}~\frac{\partial
_{x_{i}^{k}}~h_{00}}{2}= \\
&  \\
=-\frac{d~x_{i}^{k}}{d~\lambda _{i}}\frac{1}{\tilde{L}(x_{i})}\frac{d{\tilde{%
L}(x_{i})}}{d\lambda _{i}} &
\end{array}%
\end{equation}
where $b^{\prime }(x^{0})\equiv db(x^{0})/dx^{0}$. Since in the limit (\ref%
{weak field slow motion appr}) one has%
\begin{equation}
\begin{array}{ll}
\tilde{L}(x_{i})\simeq \left\vert b(x_{i}^{0})\,\frac{d~x_{i}^{0}}{d~\lambda
_{i}}\right\vert \,\Rightarrow \,\frac{1}{\tilde{L}(x_{i})}\frac{d{\tilde{L}%
(x_{i})}}{d\lambda _{i}}\simeq \left( \frac{d~x_{i}^{0}}{d~\lambda _{i}}%
\right) ^{-1}\left( \frac{b^{\prime }(x_{i}^{0})}{b(x_{i}^{0})}\left( \frac{%
d~x_{i}^{0}}{d~\lambda _{i}}\right) ^{2}+\frac{d^{2}~x_{i}^{0}}{d~\lambda
_{i}^{2}}\right) \,, &
\end{array}%
\end{equation}%
it is easy to recognize that (\ref{newtonian limit x 1}) is an identity, as
expected. On the other hand (\ref{newtonian limit x 2}) becomes%
\begin{equation}
\begin{array}{ll}
-\frac{d^{2}~x_{i}^{k}}{d~\lambda _{i}^{2}}+b^{2}\left( x_{i}^{0}\right)
~\left( \frac{d~x_{i}^{0}}{d~\lambda _{i}}\right) ^{2}~\frac{\partial
_{x_{i}^{k}}~h_{00}}{2}= \\
&  \\
=-\frac{d~x_{i}^{k}}{d~\lambda _{i}}\left( \frac{d~x_{i}^{0}}{d~\lambda _{i}}%
\right) ^{-1}\left( \frac{b^{\prime }(x_{i}^{0})}{b(x_{i}^{0})}\left( \frac{%
d~x_{i}^{0}}{d~\lambda _{i}}\right) ^{2}+\frac{d^{2}~x_{i}^{0}}{d~\lambda
_{i}^{2}}\right) ~. &
\end{array}%
\end{equation}
Since $x^{0}(\lambda )$ is arbitrary, we choose $x^{0}=\lambda $ so that the
last equation becomes%
\begin{equation}
\frac{d^{2}~x_{i}^{k}}{d~\lambda _{i}^{2}}-\frac{d~x_{i}^{k}}{d~\lambda _{i}}%
\frac{b^{\prime }(x_{i}^{0})}{b(x_{i}^{0})}=-b^{2}\left( x_{i}^{0}\right) ~%
\frac{\partial _{x_{i}^{k}}~h_{00}\left( y_{i}(x_{i})\right) }{2}~.
\label{newtonian limit x 4}
\end{equation}

We already know that the geodesics of the metric (\ref{metric x}), which now
is in the form (\ref{metric x newtonian limit}), are related to those of the
metric (\ref{metric y}) through the relations (\ref{geodesics x}-\ref%
{geodesics y}). Thus, it is easy to verify that the solutions of (\ref%
{newtonian limit x 4}) are given by%
\begin{equation}
x_{i}^{k}(x_{i}^{0})=y_{i}^{k}\left( \tau \left( x_{i}^{0}\right) \right) \,,
\label{newton equation x}
\end{equation}%
where we recall that $\tau \left( x^{0}\right) \equiv \int b(x^{0})dx^{0}$
(see (\ref{4})), provided that $y_{i}^{k}(y_{i}^{0})$ is a solution of
equations (\ref{newton equation y}). Again, we set $x_{i}^{0}=\lambda _{i}=t$%
, so that (\ref{newtonian limit x 4}) reads
\begin{equation}
\frac{d^{2}~x_{i}^{k}}{dt^{2}}-\frac{d~x_{i}^{k}}{dt}\frac{b^{\prime }(t)}{%
b(t)}=-b^{2}\left( t\right) ~\frac{\partial _{x_{i}^{k}}~h_{00}\left(
y_{i}(x_{i})\right) }{2}~\,,
\end{equation}%
while the isochronous trajectories (\ref{newton equation x}) of the
point-like masses becomes

\begin{equation}
x_{i}^{k}(t)=y_{i}^{k}\left( \tau \left( t\right) \right) \text{\thinspace }
\label{newton equation xx}
\end{equation}

Therefore the dynamics of $N$ point-like bodies in an isochronous spacetime
in the Newtonian limit is given by (\ref{newton equation xx}), so that it is
itself isochronous, as expected. More precisely, the dynamics given by the
isochronous metric (\ref{metric x}) in the Newtonian limit (\ref{metric x
newtonian limit}) corresponds to the isochronous realization of the
Newtonian $N$-body problem (\ref{newton equation y}-\ref{newton equation phi}%
) by means of the fictitious change of coordinate (\ref{repar}), according
to the standard procedure introduced in \cite{CL2007,calogero} in the
context of non-relativistic dynamical systems.

\section{Final remarks}

\label{conclusions}

As in the case of the treatments of non-relativistic dynamical
systems, including $N$-body problems with Newtonian equations of
motion ("accelerations equal forces") \cite{CL2007,calogero}, the
findings we recently reported \cite{isochronous
cosmologies,isochronous spacetimes} have an unpalatable
connotation: they indicate that the hope to ascertain which is
\textit{the} correct description of our cosmos is doomed by the
possibility to identify quite different cosmologies---for instance
displaying or not displaying the peculiar feature to be
isochronous, or featuring the effects of a Big Bang without
actually experiencing it---which are however indistinguishable by
any, reasonably conceivable, human experiment \cite{isochronous
cosmologies,isochronous spacetimes}. The unpleasant feeling caused
by this notion might perhaps resemble that experienced by
physicists educated before the discovery of quantum mechanics,
when they were confronted by the notion that it is actually
impossible to measure simultaneously with unlimited accuracy the
position and the velocity of a moving particle. However the
recognition that the microworld is governed by quantum rather than
classical mechanics provided enormous payoffs, while the
recognition of our inability to ever being able to determine what
is the correct description of our cosmos does not seem likely to
open the way to interesting developments in our understanding of
the make-up of the Universe. On the other hand it is not in
the scientific ethos to hide unpalatable theoretical developments: \textit{%
amicus Plato, sed magis amica veritas }must remain our motto.

In this spirit, we considered worthwhile to pursue these investigations by
providing a critical review of our previous findings relevant in a
cosmological context \cite{isochronous cosmologies,isochronous spacetimes}
and by ascertaining their connection with the analogous findings valid for
Hamiltonian systems and in particular for classical many-body problems
characterized by equations of motion of Newtonian type ("accelerations equal
forces") \cite{CL2007,calogero}. The results---as reported above---are not
surprising, but we nevertheless hope that they will be found of interest by
members of both communities, those focussed on general relativity and
cosmology as well as those focussed on Hamiltonian systems and classical
mechanics.

\end{document}